# Will Break for Productivity: Generalized Symptoms of Cognitive Depletion

**Lyndsey Franklin, Nathan Hodas**
Pacific Northwest National Laboratory
Richland, WA, USA
lyndsey.franklin@pnnl.gov
nathan.hodas@pnnl.gov

**Kristina Lerman**
Information Sciences Institute
University of Southern California
Marina Del Rey, CA, USA
lerman@isi.edu

**ABSTRACT**
In this work, we address the symptoms of cognitive depletion as they relate to generalized knowledge workers. We unify previous findings within a single analytical model of cognitive depletion. Our purpose is to develop a model that will help us predict when a person has reached a sufficient state of cognitive depletion such that taking a break or some other restorative action will benefit both his/her own well-being and the quality of his/her performance. We provide a definition of each symptom in our model as well as the effect it would have on a knowledge worker's ability to work productively. We discuss methods to detect each symptom that do not require self-assessment. Understanding symptoms of cognitive depletion provides the ability to support human knowledge-workers by reducing the stress involved with cognitive and work overload while maintaining or improving the quality of their performance.

**ACM Classification Keywords**
H.1.2 User/Machine Systems; H.4.1 Workflow Management; H.5.m Information Interfaces and Presentation (e.g. HCI) Miscellaneous; J.4 Social and Behavioral Sciences

**INTRODUCTION**
Evidence suggests that cognitive performance decreases following periods of sustained mental effort and that taking breaks helps knowledge workers of all types from paid-per-task Mechanical Turk workers to full-time analysts of every domain [3, 5, 21, 24, 29, 33]. For instance, [29] asserts that workers are refreshed after breaks and that workers who take breaks perform more work compared to those who do not. It has also been observed that participant resting baseline activity predicts subsequent performance declines [21]. It may be possible to improve a worker's overall performance by improving his/her baseline between periods of task activity. But there is competition between the motivation of a knowledge worker and the quality of their work. Highly motivated knowledge workers' performance may decrease in quality if they continue to work for long periods of time without some sort of break. Knowledge workers focused on completion of tasks may not be able to accurately monitor their fatigue level and may miss subtle cues indicating that they would benefit from some sort of break.

We explore the notion of cognitive depletion where knowledge workers' cognitive resources are depleted by continued work and their ability to accurately complete their task or provide meaningful analytical feedback becomes compromised. Cognitive depletion extends beyond simple task-related fatigue. Experts are particularly susceptible to cognitive depletion because they may not feel (or admit to) the normal signs of fatigue we often associate with exhaustion. Previous efforts to detect cognitive depleted states have made use of technology such as EEG, EOG, and fMRI to detect psycho-physiological changes suggesting cognitive overload [2, 5, 13, 27, 28, 34]. This works well under laboratory conditions but does not help the majority of knowledge workers in real-world settings. A better understanding of cognitive depletion and its symptoms would help knowledge workers structure their tasks and working environment to combat cognitive depletion and develop personal strategies for handling their fatigue before it affects their performance.

In this work, we identify the symptoms of cognitive depletion as they relate to generalized knowledge workers. Our purpose is to develop a model that will help us predict when a person has reached a sufficient state of cognitive depletion such that taking a break or some other restorative action will benefit both his/her own well-being and the quality of his/her work. The physiological measures of cognitive fatigue follow coherent sequences during the transition from normal states to those of high mental fatigue [5]. Others observe that cognitive depletion is predictable from behavioral cues [30]. We thus reason that these behavioral cues will follow predictable patterns and result in specific symptoms of cognitive depletion which can be used to detect its occurrence in individual knowledge workers. Cognitive depletion has distinct effects on information-processing, analytical, and executive control related behaviors [22] making it important to study separately from other attempts to study human error or the effects of physical fatigue on automatic, unthinking, or rehearsed behaviors.

The contributions of this work are as follows. First, we discuss the foundations of our perspective on cognitive depletion and the related fields from which we draw inspiration for our model. Next we discuss a generalized set of symptoms of cognitive depletion based on an extensive literature review. We provide a definition of each symptom as well as the effect it would have on a knowledge worker's ability to work productively. Finally, we discuss methods that may be used to detect each symptom without requiring self-assessment on the part of a knowledge worker.

## BACKGROUND

Our generalized symptoms model is drawn from literature covering the related fields of attention, interruption and task resumption, multitasking, error and mistake frameworks, quality control, working memory, and cognitive overload.

### Attention, Interruptions, and Task Resumption

Research into attention processes often attempts to measure and predict how long individuals can attend to a specific task. One important concept from attention research is the often observed vigilance decrement: after periods of sustained effort on a vigilance task, individuals' begin to miss cues critical to their task or workflow [2, 3, 27, 33]. Task-unrelated thoughts (TUTs) are another important phenomenon from attentional research. Individuals required to focus on a task for long periods of time often report self-distracting thoughts that are unrelated to the task at hand [1] [11]. Suppressing TUTs is a current topic of research but for our purposes, we view TUTs as a potential symptom of cognitive depletion. Interruption and task resumption research seeks to understand workflows and predict the optimal time to interrupt an individual so that the interruption is as minimally detrimental as possible (see our references for just a few examples). Such understanding is important to the study of cognitive depletion because any coping or mitigation strategies must also be minimally intrusive to workflows.

### Quality Control, Error, and Mistake Frameworks

The study of Mechanical Turk-style economies has provided a wealth of techniques for assessing the quality of worker contributions and for detecting workers who are abusing task structures for their own gain [9, 30, 31]. These techniques are beneficial to the study of cognitive depletion as they provide potential metrics that can be automatically collected without interrupting workers as well as providing insight into the working patterns of large groups of individuals. It also provides an easily accessible real-world example of an ideal scenario for cognitive depletion research: an economy based on constant completion of micro-tasks where workers are motivated to work beyond the point where their cognitively depleted state begins affecting the quality of their output.

A tremendous amount of work has been done to understand and categorize human error in a variety of contexts. This work is particularly important in fields such as air traffic control where human error endangers hundreds of lives [20, 32]. Our work and model presented here do not attempt to replicate the in depth error frameworks completed by others. Rather, we use them as a source of observations and discrete observable phenomena which may be symptoms of cognitive depletion.

### Working Memory and Cognitive Overload

Research into working memory and cognitive overload are the closest parallels to cognitive depletion that we have found and numerous works have provided the foundation for this current work. From these works important concepts such as mental and cognitive fatigue have arisen [12, 14, 22]. Such research makes important distinctions between the effects of acute workload and extended engagement. Particularly, a sudden acute workload may cause a sort of buckling stress where an individual is unable to cope while fatigue is the gradual loss of work capacity where the total amount of work accomplished in a time period is affected [14]. [22] defines mental fatigue as a change in psychophysiological state due to sustained performance. Both working memory and cognitive overload research study the immediate effects of task load in order to determine when a person is cognitively overloaded. We view cognitive overload as one mechanism which will, with extended engagement, result in a state of cognitive depletion. Highly overloaded persons will reach a state of cognitive depletion at a rate faster than those who are not cognitively overloaded. Mental fatigue as defined by [22] is then a cue that an individual is cognitively depleted. One seminal work of this field is the frequently used NASA-TLX [15] which provides a six-axis inventory for self-assessing the cognitive load of a task. Our work builds on this foundation to assess signs indicating that a person may be in a state of cognitive depletion.

## GENERALIZED SYMPTOMS OF COGNITIVE DEPLETION

We unify previous findings within a single analytical model of cognitive depletion. Our model describes a generalized set of symptoms which can be expected to appear following a period of sustained mental effort regardless of the exact task-related circumstances of a knowledge worker. It attributes the onset and progression of symptoms of fatigue to a common mechanism of cognitive depletion. This makes the model a valuable contribution for detecting cognitive depletion in a variety of situations. In the following sections we define each symptom and the sources which inspired its inclusion in our model and provide potential methods to detect the symptom that do not require direct input from an individual.

**Lack of Advancement/Progress**
A cognitively depleted person will fail to make progress on a task after a longer-than-usual amount of time [7, 25]. For example, a person playing a video game might fail to complete a level after prolonged effort. This is a within-subject effect meaning it can be detected by quantifying an individual's past activity rates on certain easily-validated tasks and then monitoring their current rate for significant deviations.

**Confusion**
A cognitively depleted person will not understand the current state of his/her task or how to complete it [7]. For example, a person attempting to solve a puzzle may not understand that they made an error several steps ago and that they must undo some work before they can resume making progress. Confusion could be detected by unnecessary repetitions, task unrelated exploration, and undoing of otherwise productive actions.

**Effort/Risk Over/Under Estimation**
A cognitively depleted person will inaccurately judge the amount or effort or risk involved in completing a task [9, 17, 32]. For example, a Mechanical Turk worker may attempt to complete a new series of micro-tasks only to discover that each is more difficult than they anticipated. This could be detected by tracking attempts at a task vs. the abandonment rate of a task (we will revisit task abandonment as its own symptom in a later section).

**Task Rushing**
A cognitively depleted person may begin attempting to complete tasks more rapidly or without usual diligence [9, 14, 20-22, 30]. For example, a person may attempt to skip steps in a well-defined process or take on less optional work. He/she might make impulsive decisions instead of properly weighing evidence. This can be detected when task completion tactics shift to automatic or satisficing techniques and may manifest as extremely rapid completion rates that are significantly faster than the rate expected from proper task management.

**Increased Decision Time**
A cognitively depleted person will be slow to make routine decisions and might seek more information to make a decision than usual [19, 21, 32]. For example, a person may take longer to complete a standard form than expected based on previous completions of the same form. This could be detected by tracking time to completion or the number of times the individual leaves and returns to the specific task. This may evolve into task rushing if a person becomes impatient through severe cognitive depletion. We will investigate such symptom progression as part of future work.

**Habituation**
A cognitively depleted person will stop responding to alerts and other attention-grabbing devices intended to direct his/her attention or prompt action [2, 3, 33]. For example, a person tasked with executing a decision when they observe specific patterns in a stream of symbols will fail to notice the evolution or significance of his/her target pattern. Habituation is similar to the vigilance decrement but includes circumstances when a person ceases to respond to unlooked-for patterns of importance. This could be detected by examining the amount of time certain alerts are given focus.

**Reaction Time Increases**
The amount of time between the appearance of a stimuli and a cognitively depleted person's response to it will increase or stagnate [21, 27, 30, 33]. For example, an air traffic controller will take a long time to respond to a critical error or a financial trader may be delayed in responding to an asset's sudden price movement. This is a straightforward measure of time between the presentation of a stimulus and the person's response, regardless of outcome.

**Data/Command Errors**
A cognitively depleted person will make more mistakes in providing input, data, commands, instructions, etc. [9, 23, 30, 32]. For example, a programmer may make an increasing number of syntax errors. Task-specific input validation will be useful for detecting this symptom. In the prior example, this would involve utilizing the compiler of the programming environment.

**Task Unrelated Thoughts**
A cognitively depleted person might experience an increase in thoughts that are not relevant to his/her overall goals [1, 11]. For example, a person working on a report may find themselves thinking about personal lunch plans instead. This symptom is particularly difficult to measure objectively and automatically. It may instead manifest as another symptom such as increased decision time, reaction time increases, spurious activity, or increased multi-tasking although eye-tracking may reveal extended gaze fixation away from a person's current task.

**Distraction**
A cognitively depleted person is less able to maintain focus in the presence of irrelevant stimuli/extraneous information [6, 10, 18, 24]. For example, someone might constantly re-open his/her email program when they see a notification of any kind. This could be tracked by comparing the amount of time spent in task-related interfaces vs. unrelated interfaces.

**Rapid Gaze/Focus Switching**
A cognitively depleted person re-focuses attention on different sub-tasks or parts of environment without interacting with any part of a significant amount of time [8, 10, 24, 30]. For example, a person reading a scholarly article may begin re-reading or skipping around without completing any section. This could be detected through focus and saccade tracking as in [25].

**Increased Multi-Tasking/Self Interruptions**
A cognitively depleted person may attempt to unnecessarily complete multiple tasks at once, increase the number of simultaneous tasks they attempt, switch between concurrent tasks more frequently, or make mid-progress switches between tasks more often [1, 10, 24]. For example, a person writing a report may begin switching between his/her report and an instant messaging program with greater frequency. This can be tracked by the number of program or interface changes a person makes for an increasing rate.

**Inattention**
A cognitively depleted person might fail to notice changes in his/her environment, side effects of his/her actions, or mistakes he/she makes, etc. [3, 20, 23, 33]. This symptom will manifest in ways similar to reaction time increases and habituation but extends to situations besides alerts and error conditions. Additionally, the number of attempts to submit invalid data before addressing errors might be tracked in cases where validation of input is possible.

**Forgetting**
A cognitively depleted person will experience a decrease in short-term/working memory capacity and would spend more time re-reading information, re-familiarizing, reorienting, etc. within the same task context [6, 18, 32]. For example, a person reading a scholarly article may re-read a section multiple times before moving on only to return and re-read the section again. Similar to rapid gaze/focus switching, this could be detected with focus and saccade tracking or by an increased use of memory-assistance tools such as a notepad program with increasingly frequent copy-paste actions.

**Increased Negative Affect**
A cognitively depleted person will use more negative-affect terms, express an increasingly negative aspect, or stop expressing positive affect [1, 4, 8, 14, 20-22, 26]. For example, someone completing a series of micro-tasks may start complaining about the task difficulty. This could be detected through sentiment or affect analysis of communication during task completion.

**Increased Task Abandonment**
A cognitively depleted person might begin to abandon tasks mid-completion [1, 7, 20]. For example, a person trying to solve a puzzle may quit the interface with the puzzle unsolved and not return to it for an extended period of time. This can be detected by tracking the rate at which applications are quit without completion of active tasks.

**Information Inventory Control Failure**
A cognitively depleted person's strategy for handling pushed information (emails, updates, text messages, phone calls, etc.) will fail and person the will become unable to continue tasks [10, 14, 19, 24]. [19] discusses how cognitively overloaded individuals cannot account for all the information pushed to them. Overload induces cognitive depletion so this symptom is an achieved state of total failure in a person's coping mechanisms. It could be detected by a person actively turning off available sources of information or attempting to disarm alerts in an environment. It could also be detected by a reduction in the person's typical rate for replying to requests for information or input.

**Strategy Inefficiency**
A cognitively depleted person will fail to form strategies for completing tasks, fail to adapt strategies to task changes, and cannot prioritize sub tasks or information needed for completing goals [1, 7, 8, 17, 20, 22, 24, 32]. For example, a person may fixate on a particular strategy for task completion even when it repeatedly fails instead of switching to a new strategy. Alternatively, the person may deviate from normally successful strategies into more impulsive or unstructured patterns of behavior. This could be detected by tracking attempted actions and their success, by a lack of diversity in information gathered, or a sudden increase in the number of applications or windows the person utilizes.

**Physical Effects**
Cognitive depletion will also manifest with physical symptoms such as drowsiness, exhaustion, clumsiness, subjective reports of fatigue, etc. [1, 5, 13, 19, 20, 21, 28]. Physical effects have a wide range of psychophysiological techniques to measure them but can require interpretation. [5] makes a distinction between drowsiness and cognitive fatigue by observing that drowsiness can fluctuate rapidly over a period of a few seconds and is made worse by rest or inactivity while cognitive fatigue would be alleviated by a restful break.

**CONCLUSION**
In this work we present a model of symptoms for detecting cognitive depletion. We build this model on the foundations of related fields such as attention, multi-tasking, human-error frameworks, and cognitive overload. Our model unifies the hitherto disparate set of symptoms by attributing them to a common mechanism. The model provides the framework for studying cognitive depletion under a variety of circumstances across multiple domains. We also contribute a starting set of objective metrics that can be used to detect these symptoms without requiring direct input or self-assessment on the part of individuals. Our future work will involve verification and validation of this model with user studies. Our initial approach will be case studies involving the interview and observations of knowledge workers. We expect our model to be refined as a result of these case studies with some symptoms being added, removed or merged. The benefits of our work towards understanding cognitive depletion include the ability to support knowledge-workers by reducing the stress involved with cognitive and work overload while maintaining or improving the quality of their performance simply by prompting them to take breaks at opportune times.